\documentclass{INTERSPEECH2023}

\interspeechcameraready

\usepackage{latexsym}
\usepackage{booktabs}

\usepackage[T1]{fontenc}

\usepackage[utf8]{inputenc}

\usepackage{graphicx}
\usepackage{amsfonts}
\usepackage{amssymb}
\usepackage{amsmath}

\usepackage{multicol}
\usepackage{multirow}
\usepackage{color}

\usepackage{balance}  %

\title{TokenSplit: Using Discrete Speech Representations for Direct, Refined, and Transcript-Conditioned Speech Separation and Recognition}
\name{Hakan Erdogan$^1$, Scott Wisdom$^1$, Xuankai Chang$^{2*}$\thanks{$^*$Work done during an internship at Google.},
Zal\'an Borsos$^1$, Marco Tagliasacchi$^1$, Neil Zeghidour$^1$,
John R. Hershey$^1$}
\address{
  $^1$Google Research\\
  $^2$Carnegie Mellon University
  }
\email{\{hakanerdogan, scottwisdom,
zborsos, mtagliasacchi, neilz, johnhershey\}@google.com, xuankaic@andrew.cmu.edu}

\begin{document}

\maketitle
 
\begin{abstract}
We present TokenSplit, a speech separation model that acts on discrete token sequences. The model is trained on multiple tasks simultaneously: separate and transcribe each speech source, and generate speech from text. The model operates on transcripts and audio token sequences and achieves multiple tasks through masking of inputs. The model is a sequence-to-sequence encoder-decoder model that uses the Transformer architecture. We also present a ``refinement'' version of the model that predicts enhanced audio tokens from the audio tokens of speech separated by a conventional separation model. Using both objective metrics and subjective MUSHRA listening tests, we show that our model achieves excellent performance in terms of separation, both with or without transcript conditioning. We also measure the automatic speech recognition (ASR) performance and provide audio samples of speech synthesis to demonstrate the additional utility of our model.

\end{abstract}
\noindent\textbf{Index Terms}: speech separation, speech recognition, speech synthesis, discrete representations, language modeling

\section{Introduction}
\label{sec:intro}
Sound separation is essential for audio scene understanding and manipulation. Speech separation is a sub-task of sound separation which has its own importance.
It aims to isolate different overlapping speech sources in a mixture,
such as in a recording of a meeting.
Many methods have been proposed recently that typically involve training a neural network on artificial mixtures of isolated speech to estimate individual sources
\cite{hershey2016deep, yu2017permutation, luo2019conv}. 
To address the lack of labeled data in real environments, 
unsupervised training methods \cite{wisdom2020unsupervised,wisdom2021sparse,tzinis2022remixit} have been proposed. Significant progress has been achieved recently; however, it is well-known that single-channel speech separation models can introduce unwanted distortions that impact perceptual quality and may be harmful for downstream tasks like speech recognition \cite{menne2019investigation,chen2018building,iwamoto2022bad}. 

In this work, we propose a novel speech separation model that uses discrete token sequences and has additional capabilities. Our method uses three kinds of discrete tokens: (1) SoundStream \cite{zeghidour2021soundstream} tokens as acoustic tokens, (2) w2v-BERT \cite{chung2021w2v} tokens as semantic tokens, (3) transcripts represented as character sequences.  %
Our model is based on a sequence-to-sequence encoder-decoder architecture. 
The semantic and acoustic tokens of the speech mixture as well as transcripts of each source in the mixture are provided at the input (see left panel of Figure \ref{fig:model}).
As the target, the model tries to predict transcripts, semantic and acoustic tokens of each source. During training, one or more kinds of input tokens are probabilistically replaced with mask tokens to simulate different modes of operation. After training, the model can be used to do speech separation and recognition given a speech mixture. The model can also do separation conditioned on one or both of the source transcripts, and speech synthesis when only transcripts are provided.
Separation performance of the model is additionally improved using ``refinement'', where the output of a conventional audio separation model is refined by token-based autoregressive generation. 

To evaluate our model, experiments are carried out on the Libri2Mix dataset \cite{cosentino2020librimix}.
The results show that our model can achieve excellent performance in terms of separation quality supported by subjective listening tests, both with and without transcript conditioning. 
As a by-product, TokenSplit also produces the ASR transcripts corresponding to both speakers. In our evaluation we observe that the quality of the predicted transcripts is not at the same level as a dedicated ASR model operating on the separated audio. In this paper we focus on separation performance, but we plan to improve ASR quality in future work.

\section{Relation to Prior Work}
\label{sec:prior}

Generative approaches for speech enhancement and speech separation \cite{maiti2019parametric,liu2019speech,shi2021discretization} consist of an encoder and separator that predicts intermediate representations of individual signals, from which separated signals can be generated by a vocoder.
Notably, \cite{shi2021discretization} uses a discrete token representation.
Such approaches can generate separated speech without interference. However, the quality of generative approaches is hampered by deviation from the phonetic content, speaker characteristics, or speaking style of the original speaker.

In \cite{zeghidour2021soundstream}, an end-to-end neural audio codec model was proposed, called SoundStream. The model is similar to a VQ-VAE \cite{van2017neural,razavi2019generating} and VQ-GAN \cite{esser2021taming}.
The SoundStream codec has been used in a recent framework of audio generation, called AudioLM \cite{borsos2022audiolm}. 
Semantic tokens are obtained from a w2v-BERT model \cite{chung2021w2v}, and acoustic tokens from a SoundStream \cite{zeghidour2021soundstream} encoder. SoundStream provides a deterministic codec that maps audio to discrete tokens and vice versa, avoiding the training and use of a separate vocoder unlike previous generative separation/enhancement approaches.
AudioLM was originally applied to the task of continuation of speech and piano audio, where the model generates audio that is consistent with an initial short prompt. Recently, this framework has also been applied to generating music from text descriptions \cite{agostinelli2023musiclm}, accompaniment for singing \cite{donahue2023singsong}, and text to speech (TTS) \cite{spear_tts}.

We use models introduced in previous work such as SoundStream \cite{zeghidour2021soundstream} and w2v-BERT \cite{chung2021w2v} to extract tokens. We also use ideas from AudioLM \cite{borsos2022audiolm} and apply them to the separation problem.
Our models can perform a variety of tasks, including unconditional speech separation, transcript-conditioned speech separation, ASR of overlapping speakers and multi-speaker TTS. Nevertheless, our focus is separation and transcript-conditioned separation capabilities of the model.

\section{Method}
\label{sec:method}

\begin{figure}[t]
    \centering
    \includegraphics[width=\linewidth]{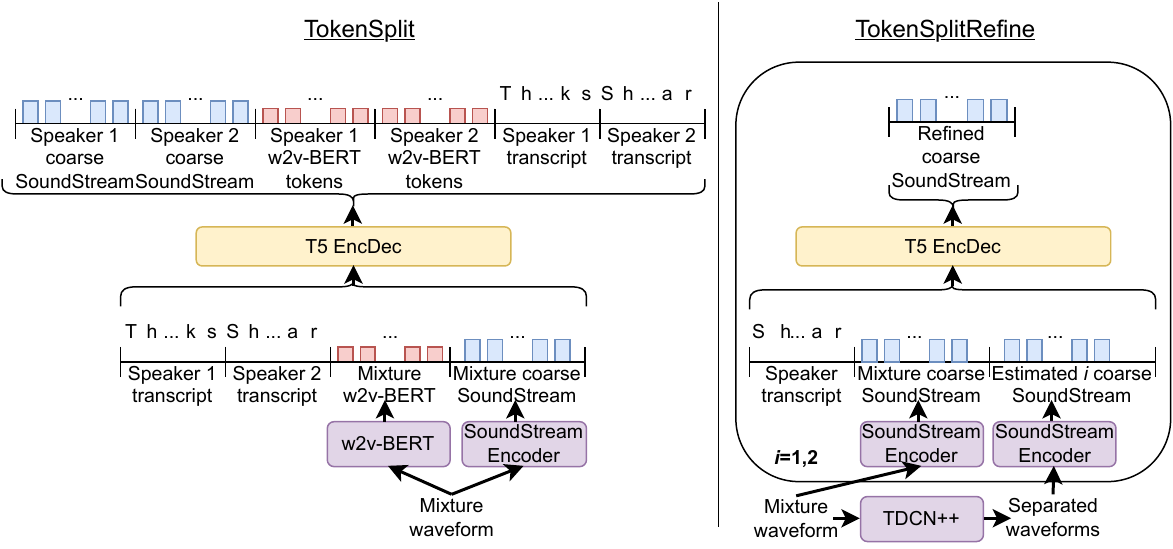}
    \caption{Illustration of models. Left: TokenSplit separation/recognition/synthesis, with optional transcript conditioning. Right: TokenSplitRefine method, where separated sources from a conventional separation model are enhanced with optional transcript conditioning. Note that TokenSplitRefine is run for each source independently, indicated by the plate notation.
    }
    \label{fig:model}
\end{figure}

In this section, we present the details of our model, as shown in Fig.~\ref{fig:model}.
We denote a mixture containing overlapping speech from $N$ speakers as $\mathbf{y} = \sum_{i=1}^{N} \mathbf{x}_i$, where $\mathbf{y} \in \mathbb{R}^T$ is the mixture of length $T$ samples and $\mathbf{x}_i \in \mathbb{R}^T$ is the $i-$th speaker's speech. In this paper, we consider the 2-speaker case, $N=2$. 

\subsection{Token Extraction}
\label{ssec:token}

Motivated by the successes of AudioLM \cite{borsos2022audiolm}, we explore using both semantic and acoustic tokens in our model. 
Since semantic tokens carry phonetic information and acoustic tokens represent the audio signal, we expect them to complement each other.

Semantic tokens are computed using w2v-BERT \cite{chung2021w2v}, which is a self-supervised learning model trained with both a masked language modeling (MLM) loss and a contrastive loss. We use a pre-trained model to map the input mixture audio waveform to embeddings extracted from an intermediate layer, which contain rich linguistic information. To discretize these embeddings, we train a \textit{k}-means model on the embeddings with $K$ clusters.
We extract the semantic tokens as follows:
\begin{align}
    \mathbf{S}_{\text{mix}} &= \text{Discretize}\left( \text{w2v-BERT}(\mathbf{y}) \right), \\
    \mathbf{S}_i &= \text{Discretize}\left( \text{w2v-BERT}(\mathbf{x}_i) \right), \  i\in\{1, \dots, N\},
\end{align}
where $\mathbf{S}_\text{mix}$ and $\mathbf{S}_i$ are semantic token sequences for input mixture and clean source $i$, respectively, and $T_S$ is the sequence length. The ``Discretize'' operation maps each embedding to the index of the closest \textit{k}-means codebook entry \cite{borsos2022audiolm}.

For acoustic token extraction, we use a SoundStream model \cite{zeghidour2021soundstream}, which is a state-of-the-art end-to-end neural audio codec at low bitrates. 
We denote the acoustic tokens as
follows:
\begin{align}
    \mathbf{A}_{\text{mix}} &=  \text{SoundStream}\left(\mathbf{y} \right), \\
    \mathbf{A}_i &= \text{SoundStream}\left(\mathbf{x}_i \right), \  i\in\{1, \dots, N\}, \label{eq:source_soundstream}
\end{align}
where
$\mathbf{A}_\text{mix}$
and
are acoustic token sequences for input overlapping speech and clean sources, with length $T_A$. The acoustic token at each time step is a vector of $Q$ dimensions.

We compute transcripts from clean signals using a pretrained ASR model instead of using ground-truth transcripts, mainly because the source speech signals are clipped to a fixed length which may chop words in the boundaries and the ground truth transcripts for them are not immediately available. We obtain
$
    \mathbf{W}_i = \text{ASR}\left(\mathbf{x}_i \right), \  i\in\{1, \dots, N\},
$
where $\mathbf{W}_i$ are encoded transcript tokens for clean sources $i$, with length $L_i$.

\subsection{Encoder-Decoder Processing}
\label{ssec:pipeline}

We use a sequence-to-sequence encoder-decoder (EncDec) Transformer architecture for our model. Cross-entropy on tokens is used as the loss function. To avoid very long sequences in the input and output,
as in AudioLM, we experimented with using two stages to first predict coarse, then fine, acoustic tokens. But in our experiments, we found that just a single stage predicting coarse acoustic tokens was sufficient. Thus, we only present results with a single stage, and report results with two stages in supplementary material \cite{project_website}.

The model takes the transcripts of each source, semantic tokens of overlapping speech, and coarse acoustic tokens of overlapping speech as input. The token values are made separate and unique before feeding into the network by shifting the token values to be in a certain range for each kind of token. We use a special separator token to separate different kinds of tokens. We apply random masking on each kind of input token sequence, so some of the sequences are fully replaced with special mask tokens. The flexibility to mask any combination of tokens allows us to train the model for a wide variety of tasks. Transcripts are padded to a fixed length of 150 with the same mask token when they are not fully replaced by mask tokens. We also apply a separate type of masking where we replace a contiguous subregion with mask tokens as well. The model learns the following mapping:
\begin{align}
    \left[
        \hat{\mathbf{A}}_{1:N}^{\leq Q''},
        \hat{\mathbf{S}}_{1:N},
        \hat{\mathbf{W}}_{1:N},
    \right]
    =& \text{EncDec}[\text{Mask}(\mathbf{W}_1),\cdot\cdot, \text{Mask}(\mathbf{W}_N), \nonumber \\ & \text{Mask}(\mathbf{S}_\text{mix}), \text{Mask}(\mathbf{A}_{mix}^{\leq Q'}) ],
\end{align}
where $\text{Mask}(\cdot)$ represents the masking function, $\mathbf{A}_{mix}^{\leq Q'}$ denotes the coarse acoustic tokens of noisy speech and $\hat{\mathbf{A}}_{1:N}^{\leq Q''}$ denotes the predicted coarse acoustic tokens of each source from 1 to $N$. Note that $Q''$ can be different than $Q'$.

To generate audio waveforms, the coarse acoustic tokens are transformed by the SoundStream decoder.

\subsection{Model Capabilities}
\label{ssec:capability}
In this part, we discuss the capability of our proposed model about joint separation, recognition and synthesis of overlapping speech through discrete representations. 
Consider three possible use cases to define different 
"inference modes" of the model.
\begin{enumerate}
    \item \textbf{Separation and recognition.} If the transcripts $\mathbf{W}_{1:N}$ are masked out at the input, the model can only see semantic tokens and acoustic tokens of overlapping speech. In such cases, the model performs speech separation and recognition.
    \item \textbf{Transcript-conditioned separation and recognition.} One or more of the transcripts are fed as input as well as the input semantic and acoustic tokens. In this case, the model can leverage the text information to predict acoustic tokens of each source.
    \item \textbf{Multi-speaker text-to-speech synthesis.} This is a side effect of this model where it can be used as a 2-speaker-TTS model when providing one or more of the input transcripts without any audio input and generating speech based on the transcript(s).
\end{enumerate}

\subsection{TokenSplitRefine}
\label{ssec:tokensplitrefine}

We also consider using a variant of our system, which we call {\it TokenSplitRefine}, to enhance the output of a traditional audio-only separation system, such as a TDCN++ \cite{kavalerov2019universal}. The outputs of this initial separation model are fed to a downstream TokenSplitRefine model, which can use optional transcript conditioning.
An illustration of this setup is shown in the right panel of Figure \ref{fig:model}.  We did not use semantic tokens for the refinement model since the benefit was minimal. 
We predict the tokens of an enhanced source from optional source transcript tokens, mixture acoustic tokens, and separated source tokens.
TokenSplitRefine can clean up distortion and artifacts present in the separated estimates. Note that this model is similar to some target speaker/speech extraction methods where an utterance of a target speaker is used to extract that speaker from a mixture. However, here we use an initial extraction of that speaker's speech as an input instead. This method is also similar to iterative architectures \cite{wang2021sequential} and generative post-processing \cite{schaffer2022music} used for sound separation problems.

\section{Experiments}
\label{sec:experiments}

\begin{figure*}[!ht]
    \centering
    \includegraphics[width=\linewidth]{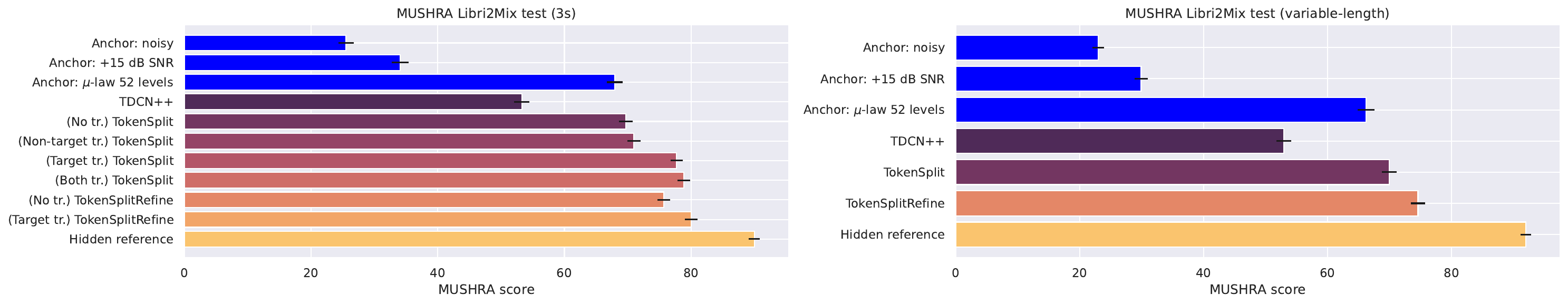}
    \vspace{-20pt}
    \caption{Results of MUSHRA listening tests on first 3s of Libri2Mix test set (left) and original variable-length Libri2Mix test (right). For input transcript conditioning, ``(No tr.)'' is none, ``(Non-target tr.)'' is non-target transcript, ``(Target tr.)'' is target source transcript, and ``(Both tr.)'' is transcripts of both sources.}
    \label{fig:mushra}
    \vspace{-13.5pt}
\end{figure*}

\subsection{Training}

TokenSplit models are trained using 10 million 3s long synthetic fully-overlapped speech mixtures which we mixed from random sources in the Libri-Light dataset \cite{kahn2020libri} with each gain sampled from ${\cal N}(-10, 4)$ dB. We used a relatively short fixed block of 3 seconds for training and testing our models since there is a limit on the token length imposed by the Transformer's limited memory. The 3s blocks were obtained from the beginning of the variable-length mixture. In addition, our models have multiple outputs, and we need to keep positional encodings consistent across multiple runs through the model which is easier to achieve with a fixed duration input and output.

To extract semantic tokens, we use w2v-BERT XL \cite{chung2021w2v} with $0.6$B parameters. Following AudioLM \cite{borsos2022audiolm}, we select the 7th layer in the MLM module of w2v-BERT XL to extract the embeddings and $K=1024$ clusters for the \textit{k}-means.
These are trained on Libri-Light \cite{kahn2020libri}, 60k hours of read English speech.

The SoundStream model used in this work uses a similar architecture as in previous work \cite{borsos2022audiolm}, however, it downsamples the 16 kHz inputs to 25 Hz embeddings. The vector quantizer has 40 layers and each layer uses a codebook size of 64. These 40 layers representation is mapped to a 20 layer representation with a 4096 codebook size for processing with the TokenSplit model. The resulting bitrate is 6000bps. We split the input acoustic token at level $Q'=10$, 
and we use levels up to $Q''=6$ for coarse target tokens.
The SoundStream model is trained on a combination of 8300 hours of Libri-Light speech mixtures and about 900 hours of LibriVox clean speech, so that it can represent both clean speech and speech mixtures well.

To obtain the transcripts used both as input and target in the model, we used a multi-lingual ASR model which has a 16 layer Conformer encoder and a 4 layer RNN decoder which was trained to recognize speech in 22 languages. 
The transcripts are encoded with character encoding and padded with mask tokens to a fixed length of 150 characters, enough to cover 3s blocks.

We use T5X \cite{roberts2022t5x} to train our proposed models. The Transformer-based encoder-decoder architecture is used in all stages, and has 267M parameters. The encoder and decoder contain 12 layers each, with 12 attention heads, embedding dimension of 768, feed-forward layer dimension of 2048 and dropout of 0.3. We used learnable absolute positional embeddings. Learnable relative position biases were added to each attention matrix as well. While training, we applied masking patterns to inputs with pre-assigned probabilities, detailed in our supplementary material \cite{project_website}. Model vocabulary included 4 special tokens, 256 transcript tokens, 1024 semantic tokens and 4096$Q$ acoustic tokens.
Training was done for 1 million steps on 16 Google Cloud TPU v4 chips with the Adafactor optimizer \cite{shazeer2018adafactor} with batch size 256 and learning rate 0.5 with 10000-step warmup and inverse square-root decay. TokenSplit trains in about 7 days, and TokenSplitRefine in about 5 days. 

Typically, separation models are trained with a permutation invariant loss function \cite{hershey2016deep,yu2017permutation}. Here, we did not do that, since teacher-forced training is used during training and autoregressive sampling is used during inference. The model is clued about the target source to decode from the previous sampled tokens. During inference, we sample the first token with model provided probabilities and this sample most likely decides which source is decoded first.

\begin{table}[t]
\centering
\caption{Results on 3s and variable-length Libri2Mix.
Column ``Transcript'' indicate which sources' transcripts are
provided at the inputs to the models.
}
\vspace{-10pt}

\resizebox{\linewidth}{!}{
\begin{tabular}{llrrrrr}
\toprule
            Model & Transcript & SI-SNRi & DNSMOS & ViSQOL &  DWER/WER &   DCER/CER \\
\midrule
\multicolumn{7}{c}{Libri2Mix test (3s)} \\
\cmidrule{1-7}
          Mixture &  -- &     0.0 &   3.39 &   1.50 &  86.6 &  62.2  \\
           TDCN++ &  -- &    12.3 &   2.96 &   2.11 &  25.1 &  14.9  \\
       TokenSplit &  No           &    -3.5 &   3.50 &   2.04 &  26.6 &  15.4  \\
       TokenSplit &  Non-target      &    -3.6 &   3.51 &   2.05 &  26.6 &  15.3  \\
       TokenSplit &  Target         &    -3.5 &   3.51 &   2.11 &  12.1 &   6.4  \\
       TokenSplit &  Both           &    -3.5 &   3.51 &   2.12 &  12.2 &   6.5  \\
 TokenSplitRefine &  No &     0.3 &   3.51 &   2.29 &  17.2 &   9.5  \\
 TokenSplitRefine &  Target    &     0.3 &   3.51 &   2.33 &  10.9 &   5.7  \\
  \multicolumn{7}{c}{Libri2Mix test (variable-length)} \\
\cmidrule{1-7}
          Mixture &  -- &     0.0 &   3.57 &   1.69 &  87.3 / 87.6 &   66.5 / 67.0 \\
           TDCN++ &  -- &    16.0 &   3.29 &   2.42 &  13.0 / 21.6 &    7.3 / 13.2 \\
       TokenSplit &  No &    -2.9 &   3.61 &   2.06 &  21.1 / 28.3 &   12.1 / 17.5 \\
 TokenSplitRefine &  No &     0.3 &   3.62 &   2.25 &  13.6 / 22.4 &    7.5 / 13.8 \\
\bottomrule
\end{tabular}
}
\vspace{-11.5pt}
\label{tab:results}
\end{table}

\subsection{Evaluation}
\label{ssec:evaluation}

We tested the models' performance on the test set of the Libri2Mix \cite{cosentino2020librimix}, sampled at 16 kHz. 
We present results on the first 3 seconds of the test set, as well as on the original variable-length test set using ``partial overlap decoding'' (section \ref{ssec:variable_length_inference}).

To measure signal quality, we use scale-invariant signal-to-noise ratio improvement (SI-SNRi) \cite{leroux2018sdr}, Virtual Speech Quality Objective Listener (ViSQOL) \cite{chinen2020visqol}, and the non-intrusive 
Deep Noise Suppression Mean Opinion Score (DNSMOS) \cite{kareddy2020dnsmos}. We also use ASR-based metrics. Since we do not always have ground-truth transcripts (e.g.\ for 3s blocks), we use differential word error rate (DWER) \cite{wang2021sequential} and differential character error rate (DCER), which use ASR on ground-truth audio as ground-truth transcript. For variable-length Libri2Mix, we also compute standard WER and CER using ground-truth transcripts. Note that since the TokenSplit(Refine) models use autoregressive sampling at their output, metrics such as SI-SNRi that rely on exact sample alignment between references and estimates are not suitable. Frame-based ViSQOL and lexical content-based DWER/WER are also not perfect, but have more tolerance for such generative sampling.

\subsection{Libri2Mix 3s Results}
\label{ssec:exp-separation}

For the first 3s of Libri2Mix test clips, we compare the separation performance of TokenSplit models to an audio-only TDCN++ model \cite{kavalerov2019universal} in the top half of Table \ref{tab:results}. This TDCN++ model was trained on the Libri2Mix training set, different from the Libri-Light-based training set we used for TokenSplit. The Libri-Light training set is larger but more mismatched to Libri2Mix,
since it contains noisier speech.

We investigated the performance of different variants of our model that use different orders of outputs and/or missing modalities, such as some variants not using semantic tokens or transcripts. We list the variants we used and their performances in our supplementary material \cite{project_website}. The best variant used the following order of output tokens: ${\bf A}_1+{\bf A}_2+{\bf S}_1+{\bf S}_2+{\bf W}_1+{\bf W}_2$ and used all inputs of ${\bf W}_1+{\bf W}_2+{\bf S}_m+{\bf A}_m$.
This variant performed the best when considering DNSMOS and DWER/DCER. The other variants generating text or semantic tokens first did not do as well in separation. The variant without semantic tokens achieved a DWER of 27.9\% compared to 26.6\% of the best variant for coarse audio, showing that semantic tokens may not be that important for this problem.

Table \ref{tab:results} also includes results for TokenSplitRefine described in section \ref{ssec:tokensplitrefine}, with or without transcript conditioning.
Notice that transcript conditioning improves performance significantly, as expected. TokenSplitRefine output is much better as compared to the TDCN++ output in terms of DNSMOS and DWER/DCER performances.

\subsection{Libri2Mix Variable-Length Results}
\label{ssec:variable_length_inference}
We performed variable-length inference through partial overlap decoding. We used 3s blocks with an overlap of 1s. The last 1s of the previous block's output is used to prompt the next block's decoding. This is possible in T5X by fixing the outputs at certain positions in the output sequence. Note that we prompted each output type (acoustic, semantic tokens) separately
by the last 1s of the previous block result, so that the prompt is not only at the beginning part of the decoding.
Transcripts were not prompted. Results are in the bottom half of Table \ref{tab:results}.

Inspecting the results, we see that the performance of TokenSplit and TokenSplitRefine are better than the TDCN++ model in terms of DNSMOS, but worse in terms of DWER and DCER. This shows that there is room for improvement for the variable-length inference case. We think this difference can be due to mismatch with 3s training data.
Another reason could be increased chance of sampling errors when generating variable-length data using partial overlap decoding.

\subsection{Listening Tests}
Since our objective metrics are imperfect (section \ref{ssec:evaluation}), we performed MUSHRA listening tests \cite{itu2014mushra} to measure the subjective quality of our models to provide a more reliable measurement of relative quality. 128 examples from the Libri2Mix test set were used, and there are 256 total examples rated since there are two sources per mixture. Each example was evaluated by 5 raters through an online interface and were required to wear headphones. We ran one listening test with variable-length audio (88 raters), and another listening test with the first 3 seconds of the same variable-length clips (97 raters).

As MUSHRA anchors, we used the input mixture, the input mixture with a 15 dB boost in SNR for the target source to simulate interference reduction, and $\mu$-law encoding with 52 levels to simulate distortion of the source. We compare the conventional TDCN++ masking-based separation model against TokenSplit and TokenSplitRefine.

Results are shown in Figure \ref{fig:mushra}. The TokenSplitRefine models are a clear winner, with their performance improving even more with transcript conditioning. All TokenSplit(Refine) models outperform the conventional TDCN++ model. Thus, our proposed model provides excellent subjective separation performance, confirming the quality measured by objective metrics. Audio demos of MUSHRA audio are provided online \cite{project_website}.

\subsection{Recognition and TTS Performance of the TokenSplit Separation Model}
\label{ssec:exp-recognition}
For ASR with our model, we measured a DWER of 71.9\% for the model's transcript predictions, which is more than double the DWER of 26.6\% for the standalone ASR model applied to audio decoded from model's coarse SoundStream token predictions.
We attribute this to conventional ASR models being trained with much large corpora that includes clean speech. In contrast, TokenSplit is only trained with much less training data that only contains mixtures.
In a sense, signal separation is an easier problem than overlapping speech recognition. TokenSplit's ASR performance can likely be improved by increasing training data, using both clean and mixed speech for training, and optimizing transcript encodings.

For TTS, we did not evaluate the output of TokenSplit using objective or subjective metrics, but informal listening (see provided audio demos online \cite{project_website}) indicates the model can generate reasonable outputs. We notice that TokenSplit randomly chooses two speakers to utter the input transcripts, and speaker identities and style remain constant throughout.

\section{Conclusion}
\label{sec:conclusion}
In this work, we introduced a multi-purpose model that can separate multiple sources from an input mixture audio with optional transcripts. Instead of using a time-domain or frequency-domain continuous-space inputs, we use discrete tokens as inputs and targets to train our model. In this novel model, we combine the semantic, acoustic and transcript tokens together as inputs and targets. A refinement version of the model takes pre-separated sources and refines them to be artifact free. Experiments on the Libri2Mix test set show that our models can achieve excellent performance compared to a conventional time-domain separation model, especially reducing artifacts and achieving high human-rated quality. Our model can also perform ASR and TTS, but more work is required to improve performance on these tasks.

\newpage
\balance
\bibliographystyle{IEEEtran}
\bibliography{mybib}

\begin{thebibliography}{10}
\providecommand{\url}[1]{#1}
\csname url@samestyle\endcsname
\providecommand{\newblock}{\relax}
\providecommand{\bibinfo}[2]{#2}
\providecommand{\BIBentrySTDinterwordspacing}{\spaceskip=0pt\relax}
\providecommand{\BIBentryALTinterwordstretchfactor}{4}
\providecommand{\BIBentryALTinterwordspacing}{\spaceskip=\fontdimen2\font plus
\BIBentryALTinterwordstretchfactor\fontdimen3\font minus
  \fontdimen4\font\relax}
\providecommand{\BIBforeignlanguage}[2]{{%
\expandafter\ifx\csname l@#1\endcsname\relax
\typeout{** WARNING: IEEEtran.bst: No hyphenation pattern has been}%
\typeout{** loaded for the language `#1'. Using the pattern for}%
\typeout{** the default language instead.}%
\else
\language=\csname l@#1\endcsname
\fi
#2}}
\providecommand{\BIBdecl}{\relax}
\BIBdecl

\bibitem{hershey2016deep}
J.~R. Hershey, Z.~Chen, J.~Le~Roux, and S.~Watanabe, ``Deep clustering:
  Discriminative embeddings for segmentation and separation,'' in \emph{Proc.
  {ICASSP}}, 2016, pp. 31--35.

\bibitem{yu2017permutation}
D.~Yu, M.~Kolb{\ae}k, Z.-H. Tan, and J.~Jensen, ``Permutation invariant
  training of deep models for speaker-independent multi-talker speech
  separation,'' in \emph{Proc. {ICASSP}}, 2017, pp. 241--245.

\bibitem{luo2019conv}
Y.~Luo and N.~Mesgarani, ``Conv-tasnet: Surpassing ideal time--frequency
  magnitude masking for speech separation,'' \emph{{IEEE/ACM} Transactions on
  Audio, Speech, and Language Processing}, vol.~27, no.~8, pp. 1256--1266,
  2019.

\bibitem{wisdom2020unsupervised}
S.~Wisdom, E.~Tzinis, H.~Erdogan, R.~Weiss, K.~Wilson, and J.~Hershey,
  ``Unsupervised sound separation using mixture invariant training,'' in
  \emph{Advances in Neural Information Processing Systems}, 2020.

\bibitem{wisdom2021sparse}
S.~Wisdom, A.~Jansen, R.~J. Weiss, H.~Erdogan, and J.~R. Hershey, ``Sparse,
  efficient, and semantic mixture invariant training: Taming in-the-wild
  unsupervised sound separation,'' in \emph{Proc. {WASPAA}}, 2021, pp. 51--55.

\bibitem{tzinis2022remixit}
E.~Tzinis, Y.~Adi, V.~K. Ithapu, B.~Xu, P.~Smaragdis, and A.~Kumar,
  ``{RemixIT}: Continual self-training of speech enhancement models via
  bootstrapped remixing,'' \emph{IEEE Journal of Selected Topics in Signal
  Processing}, 2022.

\bibitem{menne2019investigation}
T.~Menne, R.~Schl{\"u}ter, and H.~Ney, ``Investigation into joint optimization
  of single channel speech enhancement and acoustic modeling for robust
  {ASR},'' in \emph{Proc. {ICASSP}}, 2019, pp. 6660--6664.

\bibitem{chen2018building}
S.-J. Chen, A.~S. Subramanian, H.~Xu, and S.~Watanabe, ``Building
  state-of-the-art distant speech recognition using the chime-4 challenge with
  a setup of speech enhancement baseline,'' in \emph{Proc. Interspeech}, 2018,
  pp. 1571--1575.

\bibitem{iwamoto2022bad}
K.~Iwamoto, T.~Ochiai, M.~Delcroix, R.~Ikeshita, H.~Sato, S.~Araki, and
  S.~Katagiri, ``{How bad are artifacts?: Analyzing the impact of speech
  enhancement errors on ASR},'' in \emph{Proc. Interspeech}, 2022, pp.
  5418--5422.

\bibitem{zeghidour2021soundstream}
N.~Zeghidour, A.~Luebs, A.~Omran, J.~Skoglund, and M.~Tagliasacchi,
  ``{SoundStream}: An end-to-end neural audio codec,'' \emph{{IEEE/ACM}
  Transactions on Audio, Speech, and Language Processing}, vol.~30, pp.
  495--507, 2021.

\bibitem{chung2021w2v}
Y.-A. Chung, Y.~Zhang, W.~Han, C.-C. Chiu, J.~Qin, R.~Pang, and Y.~Wu,
  ``{W2v-BERT}: Combining contrastive learning and masked language modeling for
  self-supervised speech pre-training,'' in \emph{Proc. {IEEE} Automatic Speech
  Recognition and Understanding Workshop ({ASRU})}, 2021, pp. 244--250.

\bibitem{cosentino2020librimix}
J.~Cosentino, M.~Pariente, S.~Cornell, A.~Deleforge, and E.~Vincent,
  ``{LibriMix}: An open-source dataset for generalizable speech separation,''
  \emph{arXiv preprint arXiv:2005.11262}, 2020.

\bibitem{maiti2019parametric}
S.~Maiti and M.~I. Mandel, ``Parametric resynthesis with neural vocoders,'' in
  \emph{Proc. {WASPAA}}.\hskip 1em plus 0.5em minus 0.4em\relax IEEE, 2019, pp.
  303--307.

\bibitem{liu2019speech}
Q.~Liu, P.~J. Jackson, and W.~Wang, ``A speech synthesis approach for high
  quality speech separation and generation,'' \emph{{IEEE} Signal Processing
  Letters}, vol.~26, no.~12, pp. 1872--1876, 2019.

\bibitem{shi2021discretization}
J.~Shi, X.~Chang, T.~Hayashi, Y.-J. Lu, S.~Watanabe, and B.~Xu,
  ``Discretization and re-synthesis: an alternative method to solve the
  cocktail party problem,'' \emph{arXiv preprint arXiv:2112.09382}, 2021.

\bibitem{van2017neural}
A.~Van Den~Oord, O.~Vinyals \emph{et~al.}, ``Neural discrete representation
  learning,'' \emph{Advances in Neural Information Processing Systems},
  vol.~30, 2017.

\bibitem{razavi2019generating}
A.~Razavi, A.~Van~den Oord, and O.~Vinyals, ``Generating diverse high-fidelity
  images with vq-vae-2,'' \emph{Advances in Neural Information Processing
  Systems}, vol.~32, 2019.

\bibitem{esser2021taming}
P.~Esser, R.~Rombach, and B.~Ommer, ``Taming transformers for high-resolution
  image synthesis,'' in \emph{Proc. {CVPR}}, 2021, pp. 12\,873--12\,883.

\bibitem{borsos2022audiolm}
Z.~Borsos, R.~Marinier, D.~Vincent, E.~Kharitonov, O.~Pietquin, M.~Sharifi,
  O.~Teboul, D.~Grangier, M.~Tagliasacchi, and N.~Zeghidour, ``{AudioLM}: a
  language modeling approach to audio generation,'' \emph{arXiv preprint
  arXiv:2209.03143}, 2022.

\bibitem{agostinelli2023musiclm}
A.~Agostinelli, T.~I. Denk, Z.~Borsos, J.~Engel, M.~Verzetti, A.~Caillon,
  Q.~Huang, A.~Jansen, A.~Roberts, M.~Tagliasacchi \emph{et~al.}, ``{MusicLM}:
  Generating music from text,'' \emph{arXiv preprint arXiv:2301.11325}, 2023.

\bibitem{donahue2023singsong}
C.~Donahue, A.~Caillon, A.~Roberts, E.~Manilow, P.~Esling, A.~Agostinelli,
  M.~Verzetti, I.~Simon, O.~Pietquin, N.~Zeghidour \emph{et~al.}, ``{SingSong}:
  Generating musical accompaniments from singing,'' \emph{arXiv preprint
  arXiv:2301.12662}, 2023.

\bibitem{spear_tts}
E.~Kharitonov, D.~Vincent, Z.~Borsos, R.~Marinier, S.~Girgin, O.~Pietquin,
  M.~Sharifi, M.~Tagliasacchi, and N.~Zeghidour, ``Speak, read and prompt:
  High-fidelity text-to-speech with minimal supervision,'' \emph{arXiv preprint
  arXiv:2302.03540}, 2023.

\bibitem{project_website}
\emph{TokenSplit project website},
  \url{https://google-research.github.io/sound-separation/papers/tokensplit}.

\bibitem{kavalerov2019universal}
I.~Kavalerov, S.~Wisdom, H.~Erdogan, B.~Patton, K.~W. Wilson, J.~L. Roux, and
  J.~R. Hershey, ``Universal sound separation,'' in \emph{Proc. {WASPAA}},
  2019, pp. 175--179.

\bibitem{wang2021sequential}
Z.-Q. Wang, H.~Erdogan, S.~Wisdom, K.~Wilson, D.~Raj, S.~Watanabe, Z.~Chen, and
  J.~R. Hershey, ``Sequential multi-frame neural beamforming for speech
  separation and enhancement,'' in \emph{Proc. {SLT}}, 2021, pp. 905--911.

\bibitem{schaffer2022music}
N.~Schaffer, B.~Cogan, E.~Manilow, M.~Morrison, P.~Seetharaman, and B.~Pardo,
  ``Music separation enhancement with generative modeling,'' in \emph{Proc.
  {ISMIR}}, 2022.

\bibitem{kahn2020libri}
J.~Kahn, M.~Rivi{\`e}re, W.~Zheng, E.~Kharitonov, Q.~Xu, P.-E. Mazar{\'e},
  J.~Karadayi, V.~Liptchinsky, R.~Collobert, C.~Fuegen \emph{et~al.},
  ``Libri-light: A benchmark for asr with limited or no supervision,'' in
  \emph{Proc. {ICASSP}}, 2020, pp. 7669--7673.

\bibitem{roberts2022t5x}
A.~Roberts, H.~W. Chung, A.~Levskaya, G.~Mishra, J.~Bradbury, D.~Andor,
  S.~Narang, B.~Lester, C.~Gaffney, A.~Mohiuddin, C.~Hawthorne, A.~Lewkowycz,
  A.~Salcianu, M.~van Zee, J.~Austin, S.~Goodman, L.~B. Soares, H.~Hu,
  S.~Tsvyashchenko, A.~Chowdhery, J.~Bastings, J.~Bulian, X.~Garcia, J.~Ni,
  A.~Chen, K.~Kenealy, J.~H. Clark, S.~Lee, D.~Garrette, J.~Lee-Thorp,
  C.~Raffel, N.~Shazeer, M.~Ritter, M.~Bosma, A.~Passos, J.~Maitin-Shepard,
  N.~Fiedel, M.~Omernick, B.~Saeta, R.~Sepassi, A.~Spiridonov, J.~Newlan, and
  A.~Gesmundo, ``Scaling up models and data with $\texttt{t5x}$ and
  $\texttt{seqio}$,'' \emph{arXiv preprint arXiv:2203.17189}, 2022.

\bibitem{shazeer2018adafactor}
N.~Shazeer and M.~Stern, ``Adafactor: Adaptive learning rates with sublinear
  memory cost,'' in \emph{Proc. {ICML}}, 2018, pp. 4596--4604.

\bibitem{leroux2018sdr}
J.~L. Roux, S.~Wisdom, H.~Erdogan, and J.~R. Hershey, ``{SDR} - half-baked or
  well done?'' in \emph{Proc. {ICASSP}}, 2019, pp. 626--630.

\bibitem{chinen2020visqol}
M.~Chinen, F.~S. Lim, J.~Skoglund, N.~Gureev, F.~O'Gorman, and A.~Hines,
  ``Visqol v3: An open source production ready objective speech and audio
  metric,'' in \emph{Proc. {QoMEX}}, 2020, pp. 1--6.

\bibitem{kareddy2020dnsmos}
C.~K.~A.~Reddy, V.~Gopal, and R.~Cutler, ``{DNSMOS}: A non-intrusive perceptual
  objective speech quality metric to evaluate noise suppressors,'' in
  \emph{Proc. {ICASSP}}, October 2020, pp. 6493--6497.

\bibitem{itu2014mushra}
ITU, ``{Method for the subjective assessment of intermediate quality level of
  audio systems},'' \emph{BS.1534}, 2014.

\end{thebibliography}

\end{document}